\documentclass[%
 reprint,
 amsmath,amssymb,
twocolumn,
superscriptaddress
]{revtex4-2}
\usepackage{graphicx}
\usepackage{dcolumn}
\usepackage{bm}
\usepackage{color}
\usepackage{utfsym}
\usepackage{pifont}

\begin{document}
\title{Flexible manipulation of bipartite and multipartite EPR steerings}
\author{Yunyun Liang}
\affiliation{College of Physics and Electronic Engineering, Shanxi University, Taiyuan 030006, China}
\affiliation{State Key Laboratory of Quantum Optics Technologies and Devices, Shanxi University, Taiyuan
030006, China}
\author{Jing Zhang}
\email{zjj@sxu.edu.cn}
\affiliation{College of Physics and Electronic Engineering, Shanxi University, Taiyuan 030006, China}
\affiliation{State Key Laboratory of Quantum Optics Technologies and Devices, Shanxi University, Taiyuan
030006, China}
\affiliation{Collaborative Innovation Center of Extreme Optics, Shanxi University, Taiyuan, 030006, China}
\author{Rongguo Yang}
\email{yrg@sxu.edu.cn}
\affiliation{College of Physics and Electronic Engineering, Shanxi University, Taiyuan 030006, China}
\affiliation{State Key Laboratory of Quantum Optics Technologies and Devices, Shanxi University, Taiyuan
030006, China}
\affiliation{Collaborative Innovation Center of Extreme Optics, Shanxi University, Taiyuan, 030006, China}
\author{Tiancai Zhang}
\affiliation{State Key Laboratory of Quantum Optics Technologies and Devices, Shanxi University, Taiyuan
030006, China}
\affiliation{Collaborative Innovation Center of Extreme Optics, Shanxi University, Taiyuan, 030006, China}
\author{Jiangrui Gao}
\affiliation{State Key Laboratory of Quantum Optics Technologies and Devices, Shanxi University, Taiyuan
030006, China}
\affiliation{Collaborative Innovation Center of Extreme Optics, Shanxi University, Taiyuan, 030006, China}

\begin{abstract}
Bipartite and multipartite quantum steerings are significant resources for various quantum tasks, such as ultrasecure multi-user quantum network, one-site-trusted quantum communication, high-fidelity quantum computation, etc. A flexible steering manipulation scheme of five down-converted Hermitian Gaussian modes generated from an optical parametric oscillator by using a spatial structured pump is presented. In our scheme, not only the direction and types of the bipartite steering, but also different situations of multipartite steering, can be manipulated effectively, by adjusting the pump proportions with a spatial light modulator. In addition, stricter genuine pentapartite steering (only one site is trusted) can also be achieved by making the pump proportions as balanced as possible. Our scheme is versatile and experimentally feasible and offers new insights into the manipulation of steering, especially multipartite steering, which is valuable in many special quantum tasks.  
\end{abstract}
\maketitle


\section{Introduction}
Einstein-Podolsky-Rosen (EPR) steering, as a special kind of quantum correlation situated between entanglement and Bell nonlocality, was originally proposed by Schrödinger in 1935 \cite{Schrodinger1935} and formulated by Wiseman in 2007 \cite{Wiseman2007}. It characterizes the process that one party can affect the other party by performing suitable local measurements and has the feature of asymmetric quantum correlation \cite{QS1,QS2,HeQS3}, i.e., there is a direction between the involved parties and the steerabilities in two directions may not be the same. Besides bipartite steering, multipartite steering was also defined \cite{Wisema2011,He2013mult} and verified \cite{He2015mult}. Different types of quantum steerings have certain applications in different quantum information scenarios. One-way steering (only one party can steer the other) and two-way steering (both parties can steer each other) have some special applications in hierarchical quantum communication \cite{Roman2012} and high-fidelity teleportation \cite{reid2013,he2015secure}, respectively. Collective steering is helpful to build a ultra-secure multi-user quantum network \cite{He2013,Liang}. In addition, genuine multipartite steering, which was first introduced by Q. Y. He and M. D. Reid \cite{He2013mult,Reid2014criteria} and later discussed intensively by R. Y. Teh et al \cite{Teh2022}, also plays an irreplaceable role in quantum computation and quantum communication. A genuine four-partite EPR steering was experimentally realized and further applied to universal one-way quantum computing \cite{Pan2015(2)}. There are also other potential applications based on quantum steering, such as one-sided device-independent (1SDI) quantum key distribution \cite{Wiseman2012,Wiseman2016}, one-way quantum computing \cite{Pan2015(2)}, subchannel discrimination \cite{pan2015}, and randomness generation \cite{He2019randomness,passaro2015,skrzypczyk2018} etc.   

Since quantum steering has such rich application scenarios, manipulating it between its different bipartite types and among multipartite situations can provide more flexible implementation plans in quantum information. A common way to manipulate steering is making the states asymmetric by adding losses or noise to the subsystems \cite{Qin2017,Guo2017}, which is convenient and experimentally feasible to switch the steering between one-way and two-way \cite{Guo2017} or change the direction of one-way steering \cite{Qin2017,yang2021,han2022,wang2023}. However, additional loss or noise leads to a reduction in correlation and limits the secure communication distance. To avoid this limitation, Olsen proposed a scheme for controlling the asymmetry of bipartite EPR steering (from two-way to one-way) based on a seed-injected non-degenerate optical parametric oscillator, by adjusting the ratio of the injected signal to the pump amplitude \cite{Olsen2017}. Manipulation of bipartite steering asymmetry (from two-way to one-way) through interference effects induced by closed-loop coupling was also discussed \cite{He2019}. Recently, all configurations of bipartite EPR steering, including two-way steering, two one-way steerings, and the two-way unsteerable correlation, were observed simultaneously with different outputs of local filter operations on both Alice's and Bob's sides \cite{Guo2024}. 

In this paper, different from the above-mentioned bipartite steering control proposals, an effective quantum control scheme is proposed for both bipartite and multipartite EPR steerings, from a more comprehensive perspective. In our scheme, not only the types and directions of bipartite steerings but also the different situations of multipartite steering (including the genuine multipartite steering) can be manipulated by wavefront shaping \cite{Walborn} (adjusting the proportion of different components) of the pump beam. This paper is organized as follows: In Sec.~II, we introduce our system and give its Hamiltonian. In Sec.~III, we discuss the manipulation among six different types of bipartite steering by adjusting the pump proportion. In Sec.~IV, we further analyze the manipulation among different situations of tripartite and quadripartite steerings. Finally, in Sec.~V, we also consider the parameter regions of the genuine pentapartite steering. 
\section{System: an optical parametric oscillator with a spatial structured pump}
As shown in Fig.1, a structured pump beam and a weak seed beam are injected into a cavity with a type II ($\text {KTP}-\text {KTiOPO}_{4}$) crystal placed there. Then five Hermitian Gaussian (HG) modes with orthogonal polarization (details are shown in Fig.1) can be generated by the corresponding down conversion processes, i.e. $\text{HG}_{30}\to \text{HG}_{20}+\text{HG}_{10}, \text{HG}_{12}\to \text{HG}_{10}+\text{HG}_{02}, \text{HG}_{21}\to \text{HG}_{01}+\text{HG}_{20}, \text{HG}_{03}\to \text{HG}_{01}+\text{HG}_{02},  \text{HG}_{12}\to \text{HG}_{11}+\text{HG}_{01}, \text{HG}_{21}\to \text{HG}_{11}+\text{HG}_{10}$. The pump electric field can generally be written on the Hermite-Gaussian basis (HGmn) as $E_P=b\operatorname{cos}\theta \text{HG}_{30}+\sqrt{1-b^2}\operatorname{cos}\theta \text{HG}_{03}+c\operatorname{sin}\theta \text{HG}_{21}+\sqrt{1-c^2}\operatorname{sin}\theta \text{HG}_{12}$, where $b\in[0,1],c\in[0,1],\theta\in[0,\frac{\pi }{2}]$. The normalized parameters $b, c$, and $\theta$ of the pump light can be conveniently adjusted in experiments using a spatial light modulator (SLM). Specifically, $b, c$, and $\theta$ determine the relative strengths of $\text{HG}_{30}$ and $\text{HG}_{03}$, $\text{HG}_{12}$ and $\text{HG}_{21}$, ($\text{HG}_{30}$, $\text{HG}_{03}$) and ($\text{HG}_{12}$, $\text{HG}_{21}$), respectively.\\
\begin{figure}[htbp]
\centering
\includegraphics[height=3.3cm,width=8.8cm]{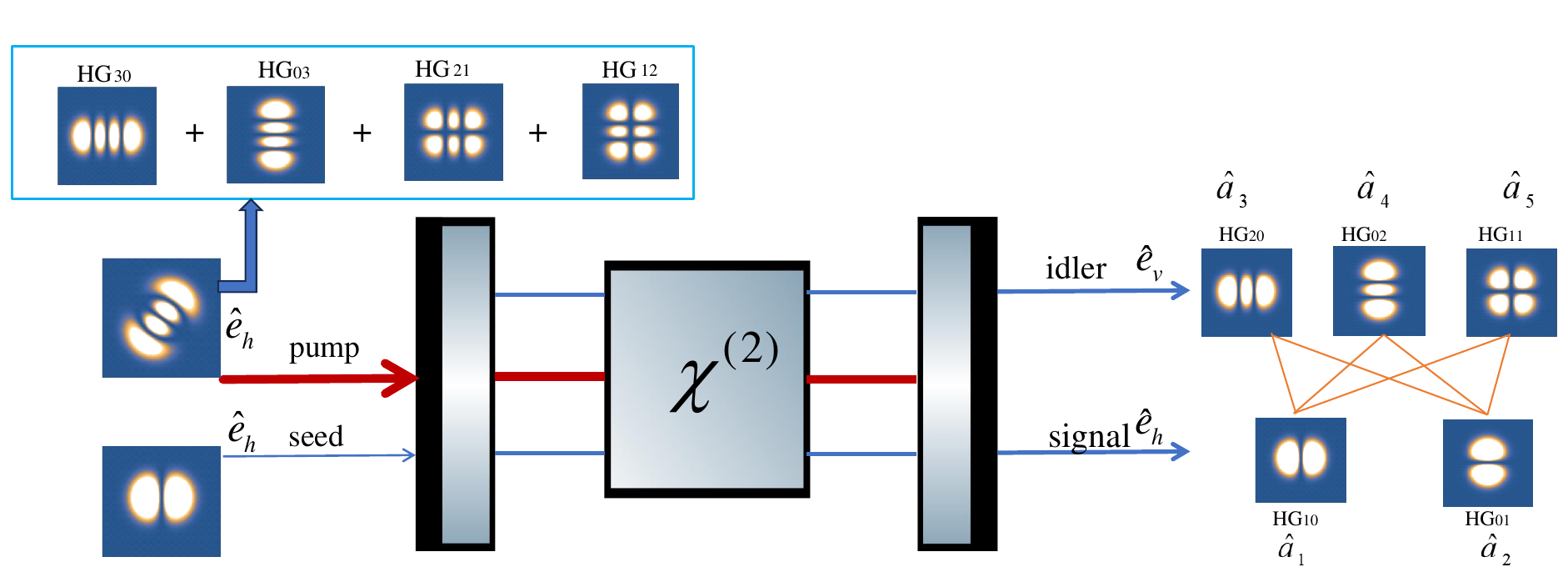} 
\caption{The type-II optical parametric oscillator injected with seed and spatial structured pump beams. Here shows an example of pump beam composed of third-order HG modes $\text{HG}_{30}$, $\text{HG}_{03}$, $\text{HG}_{21}$ and $\text{HG}_{12}$ with weights 0.222, 0.509, 0.582 and 0.593, respectively.}
\end{figure}
The Hamiltonian under the undepleted pump approximation can be expressed as
$\hat{H}=i\hbar\sum\limits_{mn}{G_{mn}\hat{a}_{m}^{\dagger }\hat{a}_{n}^{\dagger}}+\text{H.C.}$,
where $\hat{a}_{m}^{\dagger}$ ($\hat{a}_{n}^{\dagger}$) and $\hat{a}_{m}$ ($\hat{a}_{n}$) are the creation and annihilation operators of the down-converted $\text{HG}$ modes, respectively, and H. C. is the Hermitian conjugate \cite{olsen2010}. The adjacency matrix of the Hamiltonian graph, $G_{mn}={{\chi }^{(2)}} \sum\limits_{mn}{{\alpha }_l}\Lambda _{lmn}$, which depends on the second-order nonlinearity coefficient ${{\chi }^{(2)}}$, the weight of the pump ${{\alpha }_l}$, and the spatial overlap $\Lambda _{lmn}=\int{d^2r}u_l^p(r)u_m^s(r)u_n^i(r)$ with  $u_l^p(r),u_m^s(r)$ and $u_n^i(r)$ representing the pump, signal, and idler fields, respectively. The overlap integral contains the selection rules of the transverse mode coupling discussed in Refs.\cite{schwob1998,Khoury2018}. For simplicity, the down-converted modes $\text{HG}_{10}$, $\text{HG}_{01}$, $\text{HG}_{20}$, $\text{HG}_{02}$ and $\text{HG}_{11}$ are named as $\hat{a}_{1}, \hat{a}_{2}, \hat{a}_{3}, \hat{a}_{4}$ and $\hat{a}_{5}$, respectivelywhere $\hat{a}_{1}$ and $\hat{a}_{2}$ are designated as signal modes, while $\hat{a}_{3}$, $\hat{a}_{4}$ and $\hat{a}_{5}$ as idler modes. Based on the definitions of the quadrature amplitude operators ${\hat{X}_{i}}$ and the quadrature phase operators ${\hat{Y}_{i}}$, ${\hat{X}_{i}}=({{{\hat{a}}}_{i}}+\hat{a}_{i}^{\dagger } )/\sqrt{2}$ and ${\hat{Y}_{i}}=i( \hat{a}_{i}^{\dagger }-{{{\hat{a}}}_{i}})/\sqrt{2}$. The corresponding Heisenberg equations can be written in matrix form $\frac{d\xi }{dt}=M\xi$, where $\xi ={{({\hat{X}_{1}},...{\hat{X}_{5}}, {\hat{Y}_{1}}, ...,{\hat{Y}_{5}})}^T}$and the coefficient matrix $M=diag[G,-G]$. The sub-matrix $G$ can be derived by incorporating both the pump weights and the spatial overlaps. Then the Heisenberg equation can be solved by diagonalizing $M$, i.e., $M=P M_{diag} {{P}^{-1}}$, and the solution can be expressed by $\xi =S\xi (0)$, where $S$ is a symplectic matrix and $S=P{{e}^{M_{diag} t }}{{P}^{-1}}$ with $t$ representing the interaction time. $\xi (0)$ represents the input vacuum noise term from the environment through the input end 
face of the nonlinear crystal. Therefore, the covariance matrix (CM) of the five down-converted modes can be given by \cite{Loock2011,zhu2021}
\begin{gather}
\sigma =\left\langle \xi {{\xi }^{T}} \right\rangle =S \left\langle\xi (0)\xi {{(0)}^{T}} \right\rangle {{S}^{T}}=S{{S}^{T}},
\end{gather} 
from which one can obtain the quantum correlation characteristics of its related Gaussian state. It should be noted that a simplified open system is adopted here, which is one of the commonly used approaches for analyzing the OPO process with a cavity. Specifically, only vacuum noise entering through the crystal is considered, while cavity damping is neglected. This simplification focuses on the nonlinear process itself, thereby facilitating a clearer discussion of steering manipulation \cite{olsen2005,olsen2010}.
\section{The manipulation of (1+1)-steering} 
\subsection{The criterion of bipartite entanglement and steering}
For one kind of the bipartition of Gaussian state system (subsystem \textit{A} contains one mode and subsystem \textit{B} also contains one mode), the corresponding CM can be written in the form $\sigma _{AB}=\left(\begin{matrix}
   \mathcal{A} & \mathcal{C}  \\
{\mathcal{C}^T} & \mathcal{B}  \
\end{matrix} \right)$ with elements $\sigma _{AB}=\left\langle {{\xi }}_{i} {{{\xi }}}_{j} + {{\xi }}_{j} {{{\xi }}}_{i}\right\rangle/2$, where submatrices $\mathcal{A}$ and $\mathcal{B}$ are the reduced state of subsystem $A$ and $B$, respectively, and submatrix $\mathcal{C}$ corresponds to the correlation between them. Here logarithmic negativity $E_N$ is used to quantify bipartite quantum entanglement\cite{Vidal2002}, $E_N$= max $[0,-\ln (2\eta )]$, where $\eta =\sqrt{({{\varepsilon -\sqrt{{{\varepsilon }}^{2}}-4R }})/{2}}$, $\varepsilon$ = $R_1$+$R_2$-2$R_3$, $R$ = det$\sigma _{AB}$, $R_1$ = det$\mathcal{A}$, $R_2$ = det$\mathcal{B}$, $R_3$ = det$\mathcal{C}$. $E_N>0$ means that \textit{A} and \textit{B} are entangled, and the larger the $E_N$ the stronger the entanglement. While the steerability from \textit{A} to \textit{B} and from \textit{B} to \textit{A} can be defined as $ {\mathcal{G}^{A\to B}}$ = max $[0,\frac{1}{2}\ln (\frac{{{R }_1}}{4R})]$ and ${\mathcal{G}^{B\to A}}$ = max $[0,\frac{1}{2}\ln (\frac{{{R }_2}}{4R })]$, respectively. \textit{B} can be steered by \textit{A} if $\mathcal{G}^{A\to B}>0$. \textit{A} can be steered by \textit{B} if $ \mathcal{G}^{B\to A}>0$. When $\mathcal{G} ^{A\to B}>0$ and $ \mathcal{G}^{B\to A}>0$ are both satisfied, it is a two-way steering, otherwise $\mathcal{G}^{A\to B}>0$, $\mathcal{G}^{B\to A}=0$ or $\mathcal{G}^{B\to A}>0$, $\mathcal{G}^{A\to B}=0$ represents a one-way steering (only \textit{A} can steer \textit{B}, or only \textit{B} can steer \textit{A}). For two-way steerings, there are three possible conditions: symmetric two-way ($\mathcal{G}^{A\to B}= \mathcal{G}^{B\to A}$, the steerabilities from \textit{A} to \textit{B} and from \textit{B} to \textit{A} are equal), \textit{A}-dominated two-way ($\mathcal{G}^{A\to B}>\mathcal{G}^{B\to A}$, the steerability from \textit{A} to \textit{B} is bigger than that from \textit{B} to \textit{A}), \textit{B}-dominated two-way ($\mathcal{G}^{B\to A}>\mathcal{G} ^{A\to B}$, the steerability from \textit{B} to \textit{A} is bigger than that from \textit{A} to \textit{B}). By the way, $\mathcal{G}^{A\to B}=0$, $\mathcal{G} ^{B\to A}=0$ means there is no steering at all, which can be called no-way steering. 

\subsection{Manipulation of bipartite steering }

\begin{figure}[htbp]
\centering
\includegraphics[height=7.5cm,width=8cm]{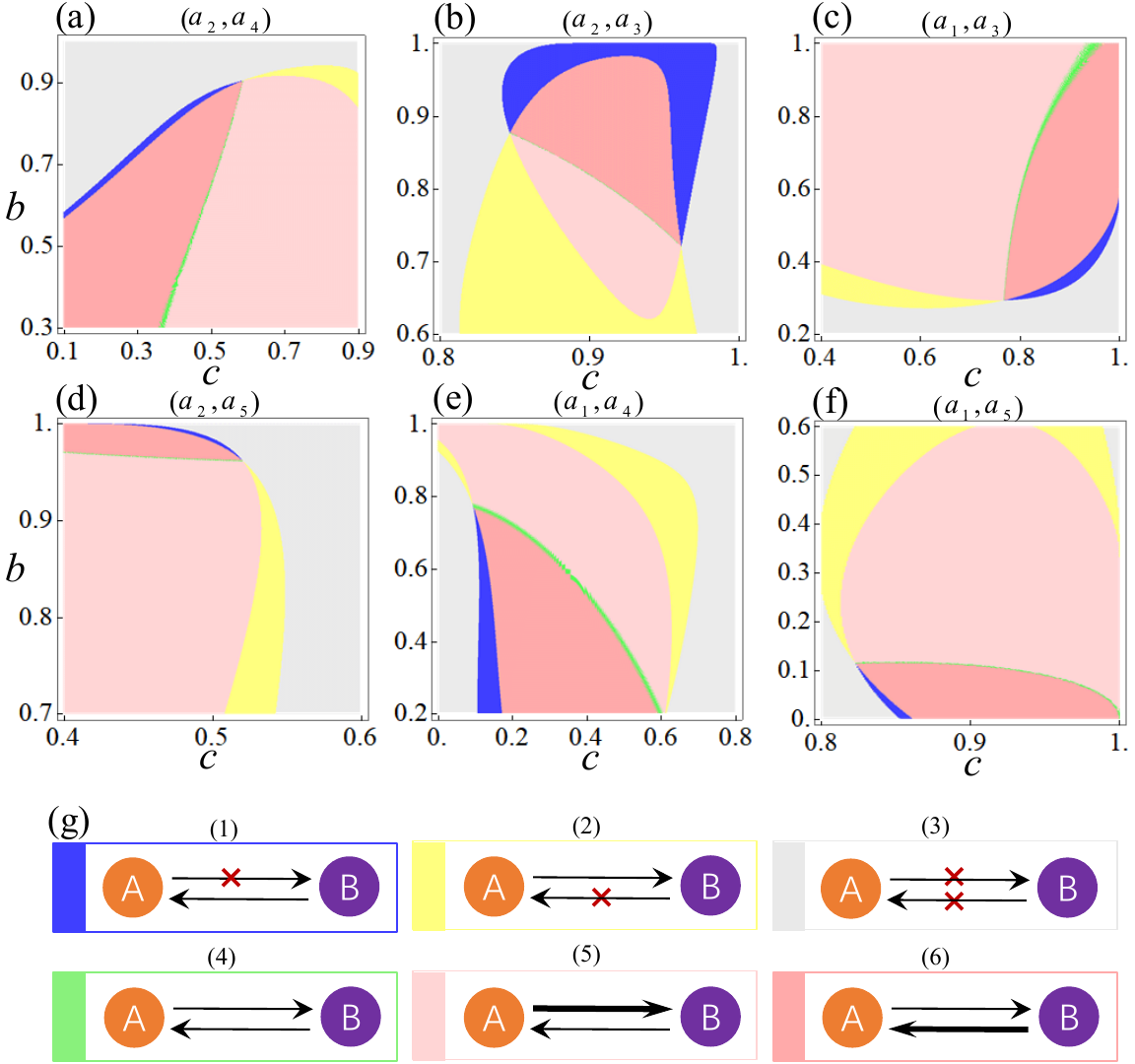}
\caption{Parameter regions of the six types of bipartite steering with $t=0.5$. The steering between $\hat{a}_2$ and $\hat{a}_4$ with $\theta=\frac{\pi }{8}$ (a), $\hat{a}_2$ and $\hat{a}_3$ with $\theta=\frac{11\pi }{32}$ (b), $\hat{a}_1$ and $\hat{a}_3$ with $\theta=\frac{\pi }{16}$(c), $\hat{a}_2$ and $\hat{a}_5$ with $\theta=\frac{3\pi }{8}$(d), $\hat{a}_1$ and $\hat{a}_4$ with $\theta=\frac{3\pi }{8}$ (e), $\hat{a}_1$ and $\hat{a}_5$ with $\theta=\frac{5\pi }{16}$ (f). (g): The legend of subgraphs (a) - (f), explaining the different colors used to represent various bipartite steering types. Here \textit{A} and \textit{B} parties correspond to the signal and idler modes, respectively.} 
\end{figure}

    For our system shown in Fig.1, all possible situations of (1+1)-steerings between \textit{A} and \textit{B} (shown in Fig.2g) are analyzed under different parameters. Sub-subgraphs (g1) and (g2) correspond to the two situations of one-way steering, $ {\mathcal{G} ^{A\to B}}=0$, ${\mathcal{G} ^{B\to A}}>0$ (blue) and $ {\mathcal{G} ^{A\to B}}>0$, ${\mathcal{G} ^{B\to A}}=0$ (yellow), respectively. Sub-subgraph (g3) corresponds to the situation of no-way steering, $ {\mathcal{G} ^{A\to B}}=0$, ${\mathcal{G} ^{B\to A}}=0$ (gray). Sub-subgraphs (g4), (g5) and (g6) correspond to the three situations of two-way steering, symmytric two-way ($ {\mathcal{G} ^{A\to B}}={\mathcal{G} ^{B\to A}}$, green), \textit{A}-dominated two-way ($ {\mathcal{G} ^{A\to B}}>{\mathcal{G} ^{B\to A}}$, light pink) and \textit{B}-dominated two-way ($ {\mathcal{G} ^{B\to A}}>{\mathcal{G} ^{A\to B}}$, pink), respectively. The regions of these six steering types for different down-converted modes are shown in Fig.2 (a)-(f), with legend (g) explaining the different colors used to represent various bipartite steering types. To better illustrate the steering manipulation, several different values of $\theta$ are selected to demonstrate all steering types and to make the parameter regime of each type as considerable as possible. Based on the configuration of the structured  pump light, $\theta$ governs the relative strength of pump ($\text{HG}_{30}$, $\text{HG}_{03}$) and pump ($\text{HG}_{12}$, $\text{HG}_{21}$). When $\theta$ is small, pump ($\text{HG}_{30}$, $\text{HG}_{03}$) dominates over pump ($\text{HG}_{12}$, $\text{HG}_{21}$), resulting in higher intensity of ($\hat{a}_1$, $\hat{a}_2$) compared to $\hat{a}_5$. Under this condition, steerability from ($\hat{a}_1$, $\hat{a}_2$) to $\hat{a}_5$ is almost maximized, while steering from $\hat{a}_5$ to ($\hat{a}_1$, $\hat{a}_2$) becomes infeasible. Therefore, selecting a large value of $\theta$ can make the steering from $\hat{a}_5$ to ($\hat{a}_1$, $\hat{a}_2$) possible, as demonstrated in Fig.2(d) and Fig.2(f). It is found in Figs.2 (a)-(f) that all possible steering situations can be obtained by controlling the parameters $b,c$ and $\theta$, for each pair of converted modes. The steerability is determined by the relative strength (in terms of mean photon number, as shown in Fig.3) between the two steering parties. The relative strengths of the down-converted modes are related to the corresponding pump weights. More involved pump beams are more likely  to achieve a stronger pump (with a wider parameter regime) than less involved pump beams. As a result, the area of the “yellow+light pink" region is larger than that of the "blue+pink" region, i.e. the steerabilities of the signal modes (three involved pumps) exceed those of the idler modes (two involved pumps) based on a comprehensive analysis. To provide a quantitative analysis, let us focus on the steering between the $\hat{a}_{2}$ and $\hat{a}_{3}$ modes as shown in Fig.2(b), where each color exhibits a considerable distribution. The corresponding involved pumps of the $\hat{a}_{2}$ and $\hat{a}_{3}$ modes are ($\text{HG}_{03}, \text{HG}_{12}, \text{HG}_{21}$) and ($\text{HG}_{30}, \text{HG}_{21}$), respectively. When considering the integrals to determine the relative strengths of the $\hat{a}_{2}$ and $\hat{a}_{3}$ modes, which itself depends on the involved pumping weights and corresponding overlap coefficients, over all parameters, we find that: 
    $I_{03}+I_{12}>I_{30}$, $I_{03}= \int_{0}^{\frac{\pi}{2}} d\theta \int_{0}^{1} db\int_{0}^{1} dc \left( \Lambda_{03,01,02}\sqrt{1-b^2}\cos\theta \right)^2=0.419$, $I_{12}=\int_{0}^{\frac{\pi}{2}} d\theta \int_{0}^{1} db\int_{0}^{1} dc \quad \left( \Lambda_{12,01,11}\sqrt{1-c^2}\sin\theta \right)^2=0.262$,  $ I_{30}=\int_{0}^{\frac{\pi}{2}} d\theta \int_{0}^{1} db\int_{0}^{1} dc \left( \Lambda_{30,10,20}b\cos\theta \right)^2=0.267$, where $ {\Lambda }_{03,01,02}, {\Lambda }_{12,01,11}, {\Lambda }_{30,10,20} $ are the overlap coefficients of the corresponding down-conversion processes ($ {\Lambda }_{03,01,02}=0.534 $, $ {\Lambda }_{12,01,11}=0.523 $, $ {\Lambda }_{30,10,20}=0.534 $). This indicates that across the entire parameter range, the intensity of the $\hat{a}_2$ mode is stronger than that of the $\hat{a}_3$ mode overall. In other words, the parameter range within which the intensity of the $\hat{a}_2$ mode exceeds that of the $\hat{a}_3$ mode is broader. When $c=0.846$, almost only one-way steering exists except for the case where $b = 0.875$ (which is a symmetric two-way steering) and the direction of the one-way steering can be manipulated by changing the parameter $b$, and it is a $\hat{a}_2\to \hat{a}_3$ one-way steering if $b<0.875$ while a $\hat{a}_3\to \hat{a}_2$ one-way steering if $b>0.875$. When $c \in (0.846, 0.961)$, it will experience $\hat{a}_2\to \hat{a}_3$ one-way, $\hat{a}_2$-dominated two-way, symmetric two-way, $\hat{a}_3$-dominated two-way, $\hat{a}_3\to \hat{a}_2$ one-way steering, sequentially, if increasing the parameter $b$. When $c=0.961$, almost only one-way steering exists again except for the case where $b = 0.725$ (which is a symmetric two-way steering) and the direction of the one-way steering can be manipulated by the parameter $b$ ($b<0.725$, $\hat{a}_2\to \hat{a}_3$ one-way steering; $b>0.725$, $\hat{a}_3\to \hat{a}_2$ one-way steering). 
\begin{figure}[htbp]
\centering
\includegraphics[height=8cm,width=8.8cm]{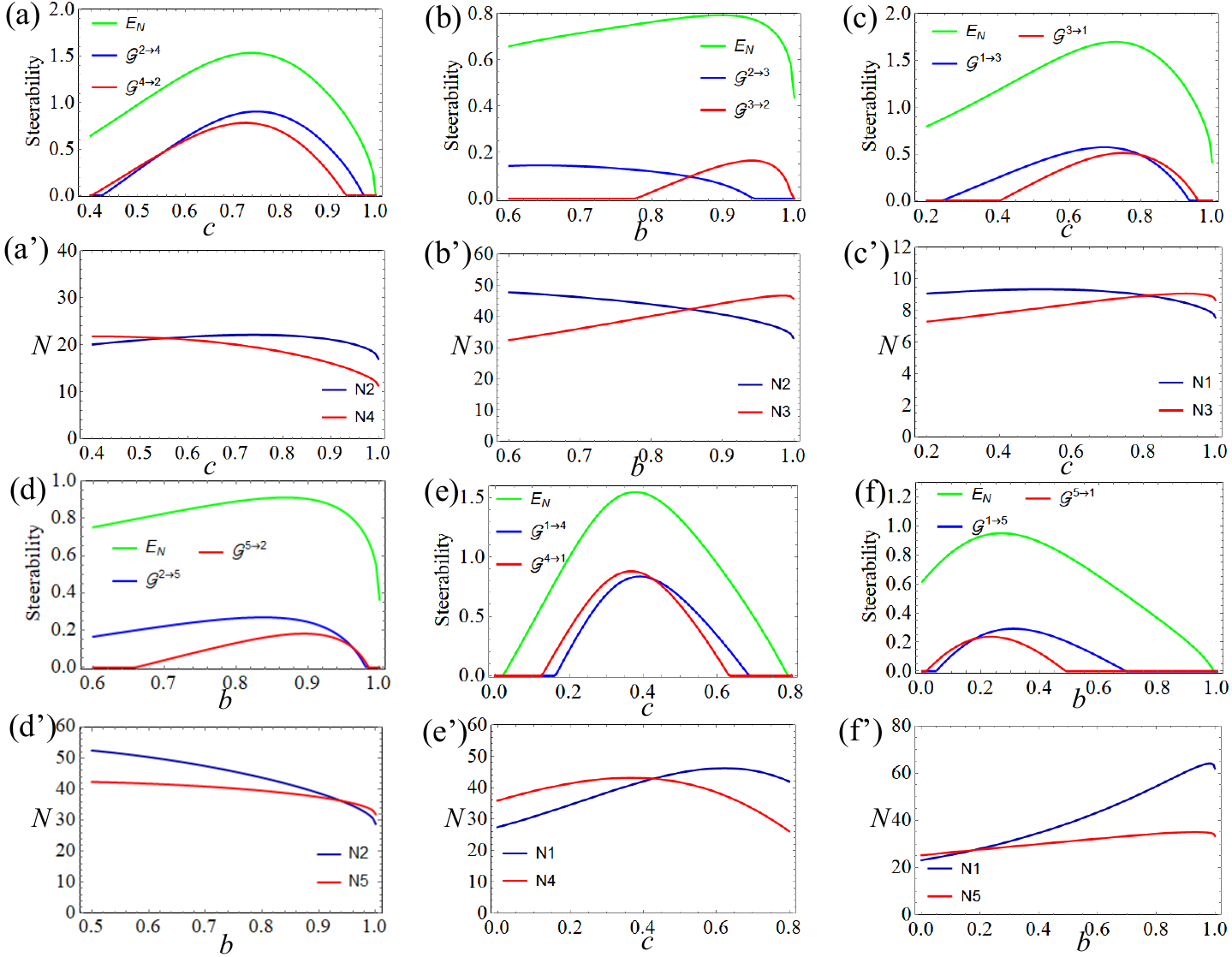}
\caption{Parameter regions of bipartite entanglement and steering with $t=0.5$ (subgraphs (a')-(f') mean photon numbers). In (a) and (a'): $\theta=\frac{\pi }{8}$, $b=0.8$; in (b) and (b'): $\theta=\frac{11\pi }{32}$, $c=0.87$; in (c) and (c'): $\theta=\frac{\pi }{16}$, $b=0.5$; in (d) and (d'): $\theta=\frac{3\pi}{8}$, $c=0.5  $; in (e) and (e'): $\theta=\frac{3\pi }{8}$, $b=0.5$; in (f) and (f'): $\theta=\frac{5\pi }{16}$, $c=0.85$.} 
\end{figure}
 Actually, for each pair of down-converted modes, it is convenient to manipulate the steering relation from $A\to{B}$ one-way, to \textit{A}-dominated two-way, symmetric two-way, \textit{B}-dominated two-way and $B\to A$ one-way steerings, in sequence, as shown in Fig.3 (a)-(f). The results of entanglement quantified by $E_N$ are also shown to verify the difference and relation between entanglement and steering, that is, steering is a subset of entanglement. These can be well understood by the behaviors of the mean photon numbers $N_A$ and $N_B$ shown in Fig.3 (a')-(f') \cite{Tan2015,He2019}. It is obvious that \textit{A}(\textit{B})-dominated steering (${\mathcal{G}^{A\to B}}>{\mathcal{G}^{B\to A}}$($ {\mathcal{G}^{B\to A}}>{\mathcal{G} ^{A\to B}}$)) corresponds to $N_A>N_B$ ($N_B>N_A$) and symmetric two-way steering (${\mathcal{G}^{B\to A}}= {\mathcal{G}^{A\to B}}$) corresponds to $N_A=N_B$.  
In other words, the steerability is determined by the relative strength of  the two steering parties, which is further related to the corresponding pump weights. Therefore, one can manipulate the (1+1)-steering flexibly among six different types shown in Fig.2 (g), by changing the proportion of four pump components, to meet with different requirements in different quantum communication tasks, especially those tasks with role changes and authority managements.

\section{Manipulation of multipartite quantum steering }
\subsection{The criterion of multipartite steering}
For the other kind of bipartition (subsystem \textit{A} contains $ n_A$ modes and subsystem \textit{B} contains $n_B$ modes), the steerability from subsystem \textit{A} to \textit{B} and subsystem \textit{B} to \textit{A} can be defined as 
\begin{equation}
    {\mathcal{G} ^{A\to B}}(\sigma _{AB})=\max \left\{ \left. 0,-\sum\limits_{j:\bar{\nu }_{j}^{\mathcal{AB/A}}<1}{\ln (\bar{\nu }_{j}^{\mathcal{AB/A}})} \right\} \right.,
\end{equation}
\begin{equation}
{\mathcal{G}^{B\to A}}(\sigma _{BA})=\max \left\{ \left. 0,-\sum\limits_{j:\bar{\nu }_{j}^{\mathcal{AB/B}}<1}{\ln (\bar{\nu }_{j}^{\mathcal{AB/B}})} \right\} \right.,
\end{equation}
where $\bar{\nu }_{j}^{\mathcal{AB/A}}(j=1,...,n_B)$ and $\bar{\nu }_{j}^{\mathcal{AB/B}}(j=1,...,n_A)$ are the symplectic eigenvalues of $\bar{\sigma} _{\mathcal{AB/A}}=\mathcal{B}-\mathcal{C}^T\mathcal{A}^{-1}\mathcal{C}$ and $\bar{\sigma} _{\mathcal{AB/B}}=\mathcal{A}-\mathcal{C}^T\mathcal{B}^{-1}\mathcal{C}$, respectively. \textit{B} can be steered by \textit{A} if $\mathcal{G}^{A\to B}>0$. \textit{A} can be steered by \textit{B} if $\mathcal{G}^{B\to A}>0$. This criterion is a sufficient and necessary condition for testing steering of Gaussian states with quadrature measurements. \cite{He2020}
\begin{figure}[htbp]
\centering
\includegraphics[height=5.4cm,width=4.5cm]{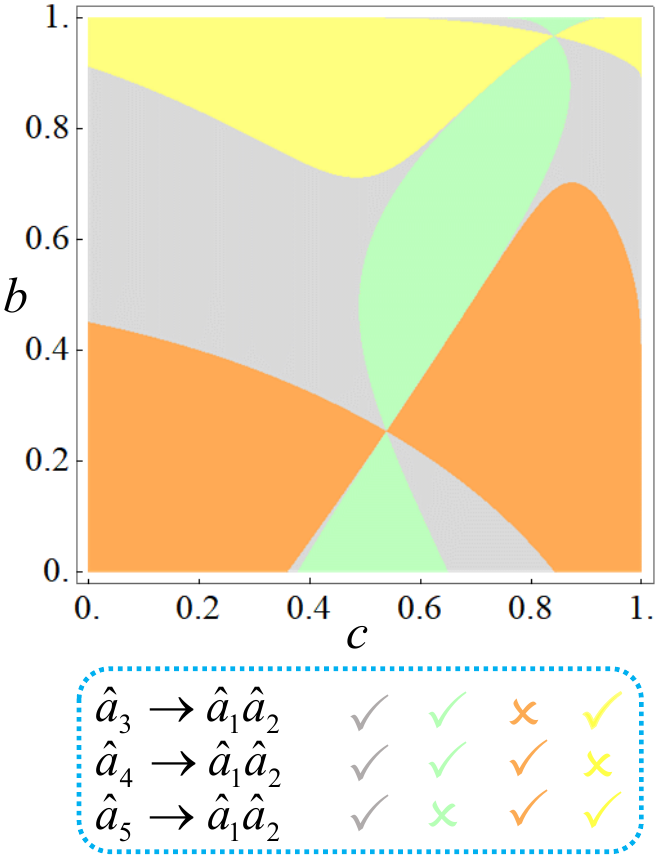}
\caption{The parameter regions of the four possible (1+2)-steerings. Here the steered part is the joint mode $\hat{a}_1\hat{a}_2$ and the steering part is $\hat{a}_3$, $\hat{a}_4$ or $\hat{a}_5$. $t=0.5, \theta=\frac{\pi }{4}$. The legend below explains the different colors used to represent various (1+2)-steering situations.}
\end{figure}
\subsection{Manipulation of tripartite quantum steering}
\begin{figure}[htbp]
\centering
\includegraphics[height=5cm,width=7.2cm]{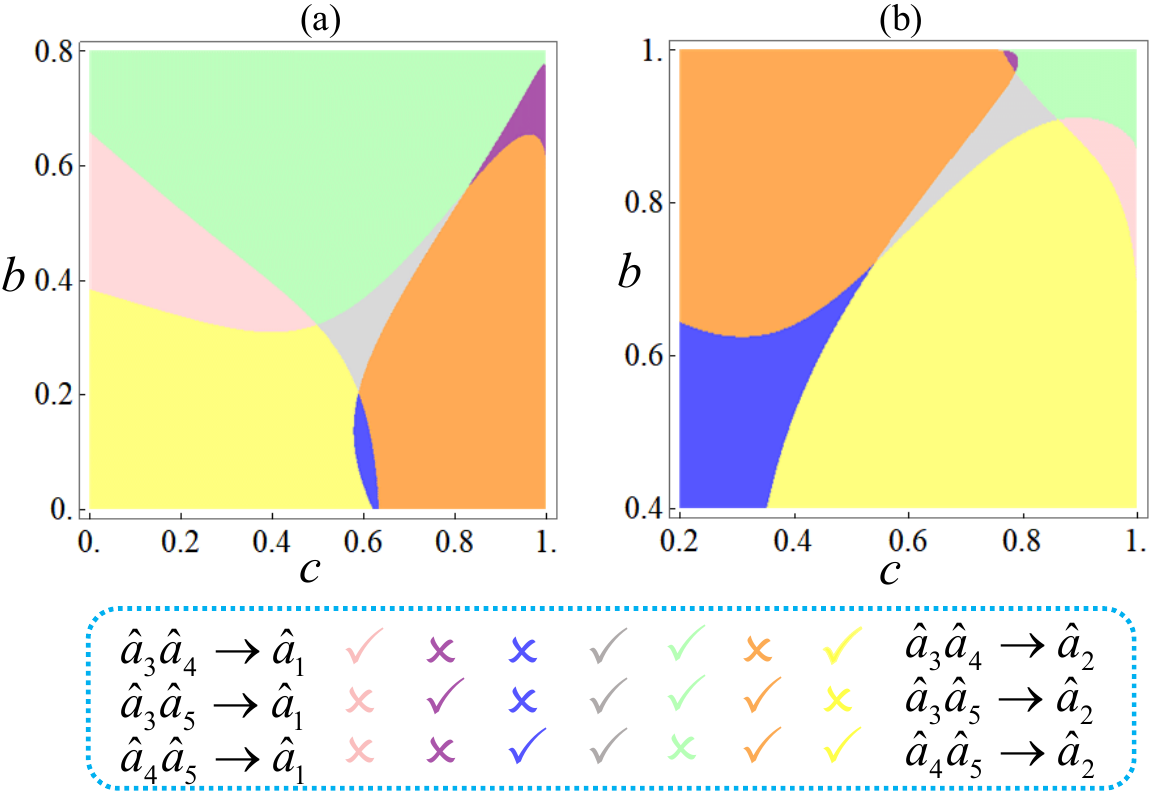}
\caption{The parameter regions of the seven possible (2+1)-steerings. Here $\hat{a}_1$ (a) or $\hat{a}_2$ (b) is chosen to be the single steered part and the joint steering part includes two of modes $\hat{a}_3$, $\hat{a}_4$ or $\hat{a}_5$. $t=0.5$, $\theta=\frac{\pi }{4}$. The legend below explains the different colors used to represent various (2+1)-steering situations.}
\end{figure}
There are eight possible (1+2)-steering situations in which the joint mode $\hat{a}_1\hat{a}_2$ can or cannot be steered by the mode $\hat{a}_3$, $\hat{a}_4$, or $\hat{a}_5$: \scalebox{0.8}{\usym{1F5F8}}\scalebox{0.7}{\usym{1F5F4}}\scalebox{0.7}{\usym{1F5F4}}, \scalebox{0.7}{\usym{1F5F4}}\scalebox{0.8}{\usym{1F5F8}}\scalebox{0.7}{\usym{1F5F4}}, \scalebox{0.7}{\usym{1F5F4}}\scalebox{0.7}{\usym{1F5F4}}\scalebox{0.8}{\usym{1F5F8}}, \scalebox{0.8}{\usym{1F5F8}}\scalebox{0.8}{\usym{1F5F8}}\scalebox{0.7}{\usym{1F5F4}}, \scalebox{0.8}{\usym{1F5F8}}\scalebox{0.7}{\usym{1F5F4}}\scalebox{0.8}{\usym{1F5F8}}, \scalebox{0.7}{\usym{1F5F4}}\scalebox{0.8}{\usym{1F5F8}}\scalebox{0.8}{\usym{1F5F8}}, \scalebox{0.8}{\usym{1F5F8}}\scalebox{0.8}{\usym{1F5F8}}\scalebox{0.8}{\usym{1F5F8}}, \scalebox{0.7}{\usym{1F5F4}}\scalebox{0.7}{\usym{1F5F4}}\scalebox{0.7}{\usym{1F5F4}}\scalebox{0.7}. All nonzero steerings among these are depicted in Fig.4, with the legend below specifying the corresponding area color for each steering situations. The steerabilities $\mathcal{G} ^{3\to 12}$, $\mathcal{G} ^{4\to 12}$, and $\mathcal{G} ^{5\to 12}$ are found to exist in parameter regions other than orange, yellow, and green, which can be understood by the related parameters of the corresponding pump modes ($\text{HG}_{30}$,$\text{HG}_{21}$), ($\text{HG}_{03}$,$\text{HG}_{12}$) and ($\text{HG}_{21}$,$\text{HG}_{12}$), respectively. Therefore, according to Fig.4(a), by changing the pump proportions, we can change the single steering part among $\hat{a}_3$, $\hat{a}_4$, and $\hat{a}_5$, with the joint steered part kept. This may be useful in multi-user quantum communication to change the single steering role of some users. By the way, other (1+2)-steerabilities $\mathcal{G} ^{1(2)\to 34},\mathcal{G} ^{1(2)\to 45},\mathcal{G} ^{1(2)\to 35}$, always exist and completely overlap, which is trivial to discuss steering manipulation. 

It is interesting that seven possible (2+1)-steering situations can exist for the steered mode $\hat{a}_1$ and $\hat{a}_2$, as shown in Fig.5(a) and Fig.5(b), respectively. The legend below gives the corresponding color for each steering situation. Here, the steering part includes two modes of $\hat{a}_3$, $\hat{a}_4$ and $\hat{a}_5$. We find that parameter regions where only one of  $\mathcal{G} ^{34\to 1(2)}$, $\mathcal{G} ^{35\to 1(2)}$, or $\mathcal{G} ^{45\to 1(2)}$ exists, and regions where two coexist, alternate around the area where all three occur. Moreover, the regions allowing only one steering are narrow, while adjacent regions allowing two to coexist are broader. This may be because filtering a specific steering is harder for multipartite steering than bipartite cases. According to Fig.5(a) and Fig.5(b), we can change the joint steering part among $\hat{a}_3$, $\hat{a}_4$, and $\hat{a}_5$, by changing the pump proportions, with the single steered part kept. This might be useful in multi-user quantum communication, to change the joint steering role of some users. 

\subsection{Manipulation of quadripartite steering}
\begin{figure}[htbp]
\centering
\includegraphics[height=3.2cm,width=8.2cm]{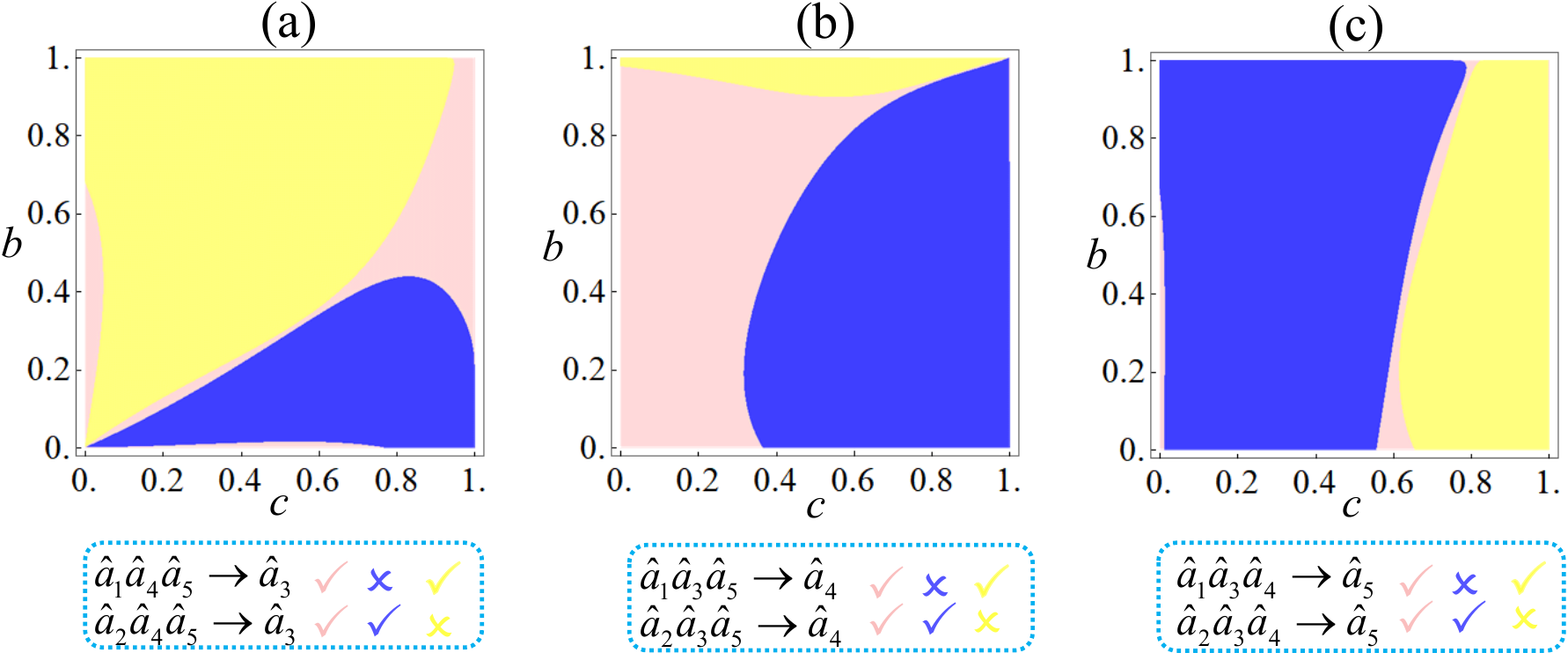}
\caption{The parameter regions of the possible (3+1)-steerings with $t=0.5$, $\theta=\frac{\pi }{4}$. Here the idler mode $\hat{a}_3$ (a), $\hat{a}_4$ (b), or $\hat{a}_5$ (c) is chosen to be the single steered part, while other two idler modes together with signal mode $\hat{a}_1$ or $\hat{a}_2$ are chosen as the steering part. The legend below explains the different colors used to represent various (3+1)-steering situations.}
\end{figure}
\begin{figure}[htbp]
\centering
\includegraphics[height=5cm,width=7.5cm]{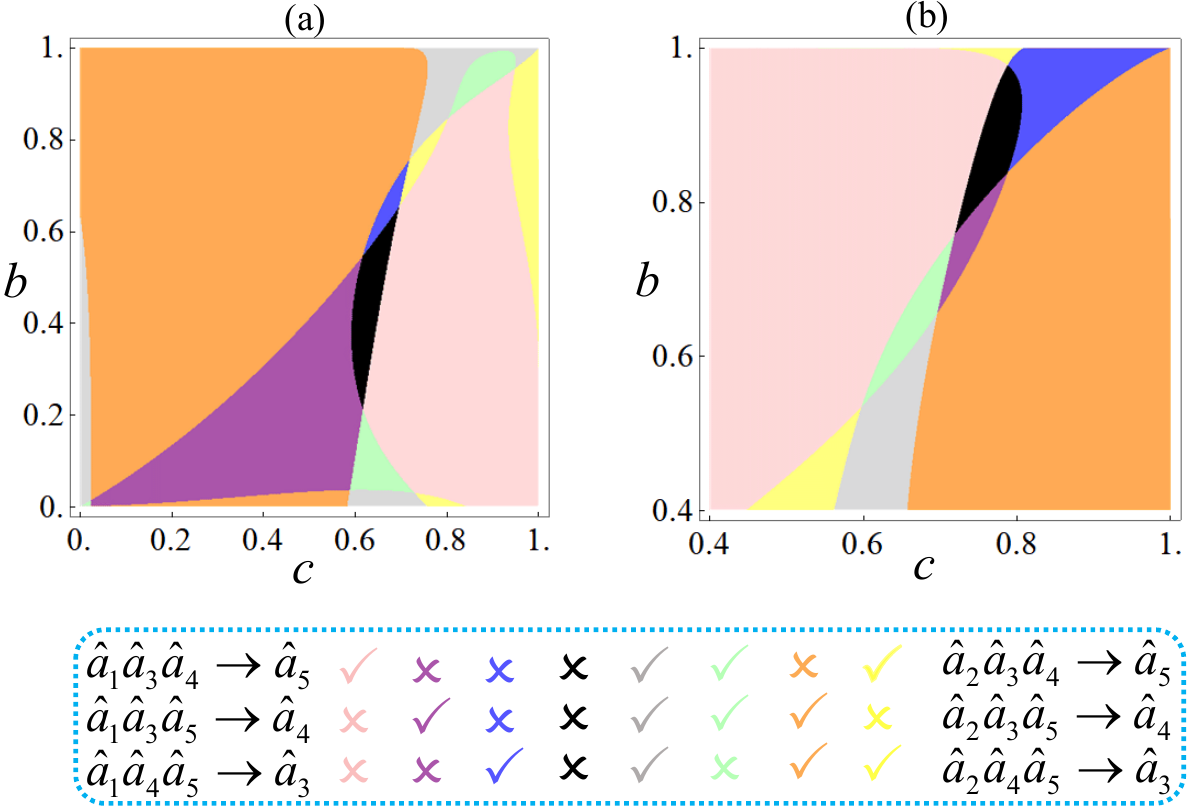}
\caption{The parameter regions of the possible (3+1)-steerings with $t=0.5,\theta=\frac{11.3\pi }{16}$. Here the signal mode $\hat{a}_1$ (a) or $\hat{a}_2$ (b) together with two of the idle modes $\hat{a}_3$, $\hat{a}_4$ and $\hat{a}_5$ are chosen as the joint steering part, while the remaining idle mode is chosen as the single steered part. The legend below explains the different colors used to represent various (3+1)-steering situations.}  
\end{figure}
The possible (3+1)-steerings with different $\theta$ values are shown in Fig.6 and Fig.7. Although there is no steering among the idler modes $\hat{a}_3,\hat{a}_4$ and $\hat{a}_5 $, two idler modes together with the signal mode $\hat{a}_1$ or $\hat{a}_2$ can steer the other idler mode separately or simultaneously. When idler modes $\hat{a}_3$, $\hat{a}_4$ and $\hat{a}_5$ are chosen as the single steered part and the other two idler modes together with the signal mode $\hat{a}_1$ or $\hat{a}_2$ as the joint steering part, the possible (3+1)-steerings are shown in Fig.6(a), Fig.6(b), and Fig.6(c), respectively. Here $\theta={\pi }/{4}$ is chosen to better demonstrate the notable outcomes of each steering situation. According to Fig.6(a), Fig.6(b), and Fig.6(c), two chosen idler modes joint with signal mode $\hat{a}_1$, $\hat{a}_2$, or both, can be manipulated by pump proportions, with the single steered part kept. This might be useful to replace some users' roles in a joint steering part with other users in ultra-safe multi-user quantum communication. It is interesting that all eight possible (3+1)-steerings can exist if we consider quadripartite steering among three idler modes and one signal mode $\hat{a}_1$, or $\hat{a}_2$, which are shown in Fig.7(a) and Fig.7(b), respectively. Here a specific $\theta$ value is chosen to better illustrate diverse steering situations. It is found that there are abundant steering situations around the diagonal line, which means that we can manipulate the (3+1)-steering among different situations by changing the pump proportion of $\text{HG}_{30}$ and $\text{HG}_{21}$ synchronously. Note that the steering part and the steered part are both changed. This may be useful in some quantum communication scenarios that the single steered user and one of the joint steering users need to exchange their roles. By the way, all (1+3)-steering situations completely overlap and out of our concern.
\section{Genuine pentapartite steering }
To meet some other requirements in quantum computation and quantum communication, such as stronger correlation and ultra-safe network, genuine multipartite steering is also under our consideration. Based on a stricter criterion \cite{Teh2022}, and considering that the standard derivation should involve two down-converted modes within the same down-conversion process as much as possible, as well as that all four pump components must be incorporated into the criteria (since the steering manipulation is dependent on adjusting their weights), the genuine pentapartite steering can be confirmed if the following inequalities are all violated, 
\begin{equation}
    \begin{aligned}
S_{I}&=\Delta (\hat{X}_3-\hat{X}_1)\Delta (\hat{Y}_1+\hat{Y}_2+\hat{Y}_3+\hat{Y}_4+\hat{Y}_5) \\&\geq   \sum\limits_{i=1}^{5}\sum\limits_{k=1,3}{{P}_{i,k}},\\
 S_{II}&=\Delta (\hat{X}_1-\hat{X}_5)\Delta (\hat{Y}_1+\hat{Y}_2+\hat{Y}_3+\hat{Y}_4+\hat{Y}_5)\\&\geq   \sum\limits_{i=1}^{5}\sum\limits_{k=1,5}{{P}_{i,k}},\\
S_{III}&=\Delta (\hat{X}_5-\hat{X}_2)\Delta (\hat{Y}_1+\hat{Y}_2+\hat{Y}_3+\hat{Y}_4+\hat{Y}_5) \\&\geq   \sum\limits_{i=1}^{5}\sum\limits_{k=2,5}{{P}_{i,k}},\\
S_{IV}&=\Delta (\hat{X}_2-\hat{X}_4)\Delta (\hat{Y}_1+\hat{Y}_2+\hat{Y}_3+\hat{Y}_4+\hat{Y}_5) \\&\geq   \sum\limits_{i=1}^{5}\sum\limits_{k=2,4}{{P}_{i,k}},\\
S_{V}&=\Delta (\hat{X}_4-\hat{X}_3)\Delta (\hat{Y}_1+\hat{Y}_2+\hat{Y}_3+\hat{Y}_4+\hat{Y}_5) \\&\geq   \sum\limits_{i=1}^{5}\sum\limits_{k=3,4}{{P}_{i,k}}.
    \end{aligned}
\end{equation} 
Here ${\Delta }$ is the standard deviation and $S_{I}$, $S_{II}$, $S_{III}$, $S_{IV}$, and $S_{V}$ represent the steering situations in which only single site, $\hat{a}_3 $ or $\hat{a}_1 $, $\hat{a}_5 $ or $\hat{a}_1 $, $\hat{a}_5 $ or $\hat{a}_2 $, $\hat{a}_2 $ or $\hat{a}_4 $, $\hat{a}_3 $ or $\hat{a}_4$, can be trusted, respectively. $P_{i,k}$ expresses the corresponding probability for a certain bipartition, with the subscript $i$ representing the chosen single site of the bipartition and the subscript $k$ denoting the trusted site. Since $\sum\limits_{i=1}^{5}\sum\limits_{k=1}^{5}{{P}_{i,k}}=1$, an equivalent criterion to confirm the genuine pentapartite steering is the violation of the inequality: 
\begin{equation}
    \begin{aligned}
S_{I}+S_{II}+S_{III}+S_{IV}+S_{V} \geq 2.
    \end{aligned}
\end{equation} 
The ($b,c$), ($\theta,b$) and ($\theta,c$) parameter regions in which genuine pentapartite EPR steering exists, bounded by the pink curve, with $t=0.2$, are shown in Fig.8(a), Fig.8(b) and Fig.8(c), respectively. It is obvious that genuine pentapartite steering can be achieved in a wide region. Moreover, optimal genuine pentapartite steering corresponds to the situation in which the proportions of the pump modes $\text{HG}_{30},\text{HG}_{03},\text{HG}_{21},\text{HG}_{12}$ are almost balanced, due to the stronger correlation among all the down-converted modes. In other words, in our scheme, one can get a strong genuine pentapartite steering by making the pump proportions as balanced as possible. To summarize, our scheme can provide a versatile quantum resource to meet many requirements in quantum communication, quantum network, and quantum computation, such as roles swapping, role changing, different trusted levels of different nodes, only single trusted site, etc., directly by manipulating the pump proportions.    
\begin{figure}[]
\centering
\includegraphics[height=2.8cm,width=9.1cm]{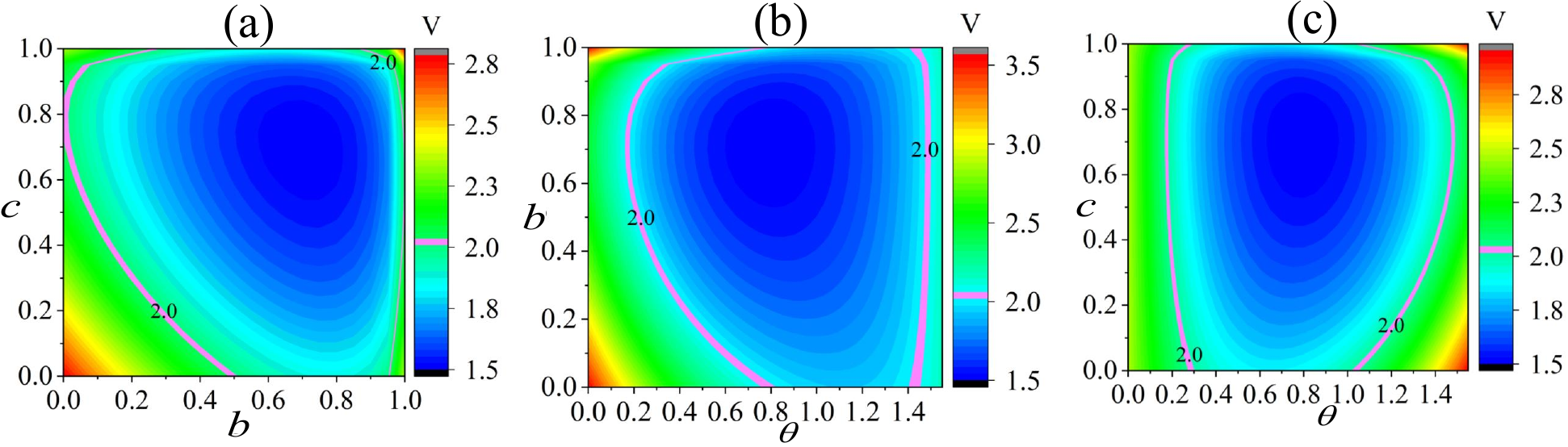}
\caption{The different parameter regions of the genuine pentapartite steering with $t=0.2$, and $\theta=\frac{\pi }{4}$ (a), $c=0.7$ (b), $b=0.7$ (c)}. 
\end{figure}

\section{Conclusions}
In summary, we propose a novel scheme to flexibly manipulate the EPR steering among five down-converted HG modes produced from a set of parametric down conversion (PDC) processes, by wave-front shaping of the pump beam. Our results show that all possible six (1+1)- steering types, most possible (1+2)/(2+1)-steering situations, all possible eight (3+1)-steering situations, and genuine pentapartite steering, can be obtained separately and switched from one another, by adjusting the weights of pump proportions (using a phase-hologram-designed spatial light modulator in experiment). Making the pump proportions balanced (unbalanced) is benefit to achieve rich steering situations (genuine multipartite steering). Note that the manipulation of multipartite EPR steering discussed here is an important issue but has rarely been investigated. Our scheme is experimentally feasible and it provides a new thought to manipulate quantum steering effectively and flexibly, which might be valuable for some special tasks in quantum communication, quantum network, and quantum computation, such as roles swapping, role changing, different trusted levels, ultra-safe multi-user network, etc.

\section{acknowledgments}
This work was supported by: Innovation Program for Quantum Science and Technology (2023ZD0300400); National Key Research and Development Program of China (Grants No. 2021YFC2201802,  2021YFA1402002); and National Key Laboratory of Radar Signal Processing (JKW202401).
\nocite{*}

\bibliography{apssamp}

\end{document}